# GERMANIUM SEGREGATION IN CVD GROWN SiGe LAYERS FOR FLASH MEMORY APPLICATION

**Andrei G. Novikau, Peter I. Gaiduk**

*Belarusian State University, prosp. Nezavisimosti, 4, 220030, Minsk, Belarus*
*E-mail: nowikow@biz.by, gaiduk@phys.au.dk*

**ABSTRACT**

A 2D layer of spherical, crystalline Ge nanodots embedded in $SiO_2$ was formed by low pressure chemical vapour deposition combined with furnace oxidation and rapid thermal annealing. The samples were characterized structurally by using transmission electron microscopy and rutherford back scattering spectrometry, as well as electrically by measuring C-V and I-V characteristics. It was found that formation of a high density Ge dots took place due to oxidation induced Ge segregation. The dots are situated in the $SiO_2$ on the average distance 5,6 nm from the substrate. Strong evidence of charge storage effect in the crystalline Ge-nanodot layer is demonstrated by the hysteresis behavior of the high-frequency C–V curves.

## 1 Introduction

Group IV elements nanocrystals (NCs-Si, Ge) embedded in $SiO_2$ have attracted much attention due to their possible applications in integrated optoelectronic devices and high density nonvolatile memories [1, 2]. It has been shown that such NCs contained in thin gate oxides exhibit charge storage properties with nonvolatile or DRAM-like memory behaviour [3, 4]. Due to the discrete charge traps (nanocrystals), the use of a floating gate composed of isolated nanodots reduces the problems of charge loss encountered in conventional flash memories. Thus, it is possible to achieve the lower operating voltages, the better endurance and faster write/erase speeds due to the use of the thinner injection oxides.

Self-assembling of silicon or germanium NCs in $SiO_2$ layers fabricated by using low-energy ion implantation and different deposition techniques has been studied by several groups [5-8]. Strong memory effects in MOS devices using oxides with such NCs were reported. However, the implantation technique process can cause Ge to be located at the silicon – tunnel oxide interface, forming trap sites that can degrade device performance [5]. The growth technique was demonstrated in the previous articles [9, 10], which is based on molecular beam epitaxialy deposition of 0,7-1 nm thick Ge layer and followed by rapid thermal processing. However, the MBE deposition can not be used for serial production yet. That's why an alternative method for production of Ge nanocrystals by using a simple and well compatible with conventional technology growth technique was proposed. The method includes the following steps: low pressure chemical vapour deposition (LPCVD) of a thin SiGe layer, thermal wet and dry oxidation, and thermal treatment in inert ambient. Therefore, the present work demonstrates the method of LPCVD formation of Ge nanocrystals for flash memory application based on oxidation induced Ge segregation.

## 2 Experiment

Before the LPCVD deposition, the uniform and pure $SiO_2$ (tunnel oxide) layer of 5 nm thickness (tunnel oxide) was grown on chemically cleaned 4 in. n-type (001) Si wafers in a dry oxygen ambiance. Then a SiGe layer of 22 nm thickness was grown on the tunnel SiO2 at 560 °C. The sandwiched SiGe/SiO2/Si samples were furnace oxidized in both wet and dry oxygen ambiance at 850 - 950 °C between 10 and 90 min. Based on the results of Fukuda et al. [11], in which the occurrence of nucleation process below the melting temperature of Ge (937 °C) and faster Ge segregation at the $Si/SiO_2$ interface for high temperature annealing was shown, we have chosen the oxidation temperature much lower (850 °C) than the Ge melting point, as well as two very close temperatures (900 °C, 950 °C). After thermal oxidation we have heated the treated samples in the inert ambient ($N_2$). A reference sample without the Ge NCs inside a 5-nm thick $SiO_2$ layer was prepared.

The structure of the samples was characterized by transmission electron microscopy (TEM) of plan-view and cross-sectional geometries with a Philips CM20 instrument operating at 200 keV. The composition and structure of the samples were controlled by Rutherford backscattering spectroscopy (RBS) using a 1.0-1.5 MeV $He^+$ beam and the spectra were analyzed by the RUMP program. Aluminium gates (area 500×500 $\mu m^2$ and separation distance of 100 μm between the neighbouring capacitors) have been fabricated for high-frequency capacitance-voltage (CV) measurements.

## 3 Results and discussion

The typical RBS spectra from as grown and thermally treated samples are shown in Fig. 1. The RBS spectra simulation shows, that Ge concentration in 22 nm thick SiGe layer is 4,5 % or 9.9

Å corresponding. It is clearly observed, that the Ge peak moves in the region of smaller energies after thermal oxidation, i.e. the back scattering of the He ions is carried out deeper and thus, it can be suggested, that we observe the Ge pile up at the $SiO_2$/SiGe interface. The profile of the Ge pileup, formed by the complete rejection of Ge atoms from the oxide during the oxidation of poly-SiGe layer, testifies the segregation process. Such oxidation behaviour can be explained successfully by a classical binary alloy oxidation theory. There are two reasons for Ge segregation during thermal oxidation in a wet or dry ambient: first, suitability of Ge is very low in $SiO_2$, lower than 0,1 % [12], and the second reason is a large difference in Gibbs pure $GeO_2$ and $SiO_2$ energies formation [13, 14]. The comparison of spectra on the inset Fig. 1 shows, that no Ge loss was found after the *dry* thermal oxidation and no Ge diffusion was found during the thermal annealing in inert ambient. However, the *wet* thermal oxidation of samples leads to the huge Ge loss (about 30 %). Ge incorporation into $SiO_2$ can be explained, if we take into account, that Ge diffusion at 850 °C is slow and the oxidation constant for wet oxidation is two magnitudes higher then the oxidation constant for a dry oxidation process. After that Ge becomes partially oxidized (forming GeO) and evaporates from the oxide.

The results of TEM investigations correlate well with RBS data. Fig. 2 represents bright- and dark-field images of a sample from the wafer oxidized at 900 °C for 15 min in O2 (the wafer is not completely oxidized). Continuous polycrystalline Ge layer is clearly observed on the bright-field cross-section image (Fig. 2a). The evidence of the polycrystalline structure is shown on Fig. 2b by presenting of a dark-field plan-view image in the Ge (111)-ring. After the whole SiGe layer oxidation and rapid thermal processing, the formation of NCs sheet takes place (Fig. 2c, 2d). Ge NCs show a dark contrast on the grey background of the $SiO_2$ layer. Spherical and well-separated Ge clusters embedded in the $SiO_2$ layer are clearly observed. Size and aerial density of Ge NCs were measured on bright-field images of a plan-view and cross-section samples from wafer, oxidized at 900 °C for 30 min in $O_2$ followed by reduction at 900 °C for 30 s in $N_2$. Typical values of size and aerial density are found to be 4 - 20 nm and $2*10^{11}$ cm$^{-2}$ respectively. The average distance of the dots from the Si substrate is measured to be 5,6 nm and the average distance between the dots can be directly measured on such XTEM images. These conditions are found to be close to optimal for the given as grown layers. The oxidation and reduction conditions were optimized with consideration of negligible Ge segregation at the Si/$SiO_2$ interface, the uniform dot-size distribution, a dot density $<10^{12}$ cm$^{-2}$ (electron transport between the dots was observed for a dot density larger than $10^{12}$ cm$^{-2}$), and the largest possible charge storage capability.

The self-assembling phenomenon of Ge nanodots in $SiO_2$ can be explained using two mechanisms. The Germanium solubility in $SiO_2$ is low and thus, the obtained after Ge segregation and piling up Ge between two $SiO_2$ layers (tunnel oxide and $SiO_2$ capping layer) structures were transformed into the supersaturated solution. It is well known that under the thermal treatment decomposition of supersaturated solution takes place, resulting in a new phase formation – Ge nanoclusters. The Ostwald ripening mechanism starts on the next stage, which describes the growth of larger particles at the expense of smaller particles. It requires diffusion of Ge atoms from the valley regions of Ge islands towards their respective centers in order to construct spherical dots for achieving greater volume to surface ratio. Thus, the Ge cluster formation takes place after the complete oxidation of the SiGe capping layer, but still during the oxidation process.

During thermal treatment in a $N_2$ ambiance, the partially oxidized Ge ($GeO_2$ clusters) is reduced by Si atoms arriving from the $SiO_2$/Si interface. It is revealed that Ge dots are formed near their initial positions, that happens due to a smaller Ge diffusion coefficient compared to Si. A reduction time longer than the optimized one leads to the formation of bigger Ge NCs by coalescence of smaller Ge NCs and to a concomitant shrinkage in the tunneling distance.

A strong evidence of charge storage effect in the crystalline Ge-nanodot layer is demonstrated by the hysteresis behaviour in the high-frequency C–V characteristic for a sample from wafer oxidized in conditions close to optimal (Fig. 3). High positive and negative gate voltages cause the C–V curves to be shifted in the direction of stored negative and positive charges, respectively. In the former case, charge trapping occurs through electron and a hole injection from the substrate into the oxide. A gradual increase in the flat-band voltage shift with increasing Vg sweep until 5 V is also observed and the voltage shift is measured to be 0.8 V for Vg 5 V. No flat-band voltage shift is observed for the reference sample prepared from wafer with pure $SiO_2$ by oxidizing at 850 °C for 60 min in $O_2$ ambiance, implying that the memory effect is Ge nanocrystals related.

One of the major methods for loss of charge in floating gate structures is leakage and subsequent loss to source and drain regions. For these reasons we measure the current-voltage curves in order to test whether the oxide between the nanocrystals can be kept large enough to suppress the direct transport between the nanocrystals. The resulting IV curves from wafers oxidized in both wet and dry ambient are shown on Fig. 4. It's observed that the dry oxidations leads for almost one magnitude lower leakage current in comparison with the wet oxidation samples. Typical leakage current for the dry oxide containing Ge NCs is measured at the level of $10^{-8}$ A/cm$^2$. We suppose, that such a promising data retention value of the leakage current was achieved due to the high quality dry thermal growth of both tunnel and capping oxides compared to the deposited oxides, used in other methods of nanocrystals MOS capacitor formation [15].

## 4      Conclusion

We have experimented with the production of thin $SiO_2$ layers with embedded Ge NCs by combination of CVD, thermal oxidation, and a thermal reduction process. By this fabrication technique it is possible to produce a sheet of crystalline Ge nanodots at any desirable depth in the oxide. We demonstrated the production of an area-dot density of $2*10^{11}$ cm$^{-2}$ of crystalline Ge dots of 4-20 nm in diameter situated in the silicon oxide 5 - 6 nm from the crystalline Si substrate. The memory effect is characterized by C-V characteristics on Al-gate capacitors, and the experiments demonstrate both hole and electron injection from the substrate. Memory windows of about 0.5 - 0.8 V for gate-voltage sweeps of 3 - 5 V are achieved.

**ACKNOWLEDGEMENTS**
This work is a part of the Belarusian Scientific Research Program "Electornics" and is funded as Electornics 1.06 projekt. We would like to acknowledge the help received from RPC "Integral" in growth SiGe samples and carrying out CV measurements.

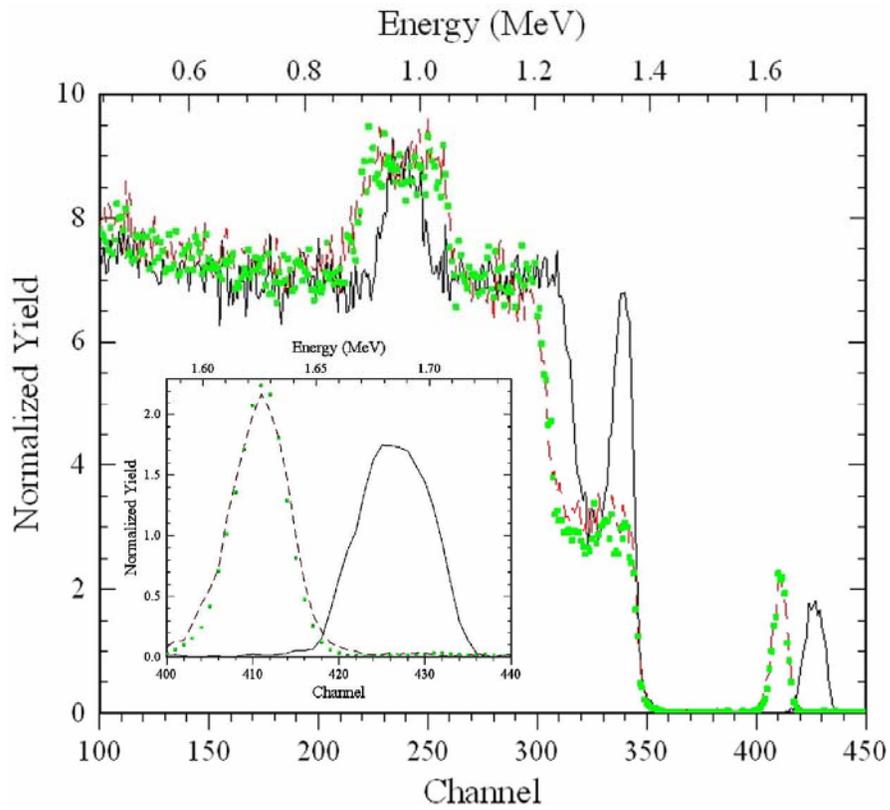

**FIGURE 1** RBS spectra from as grown SiGe/SiO$_2$/Si structure (solid line) compared to spectra from sample thermal oxidized at 900 °C for 30 min in O$_2$ (dotted line) followed by thermal annealing in N$_2$ at 900 °C for 30 s (spots) are presented. The RBS spectra behavior evidences strong Ge segregation during oxidation process. No Ge redistribution after thermal annealing is found.

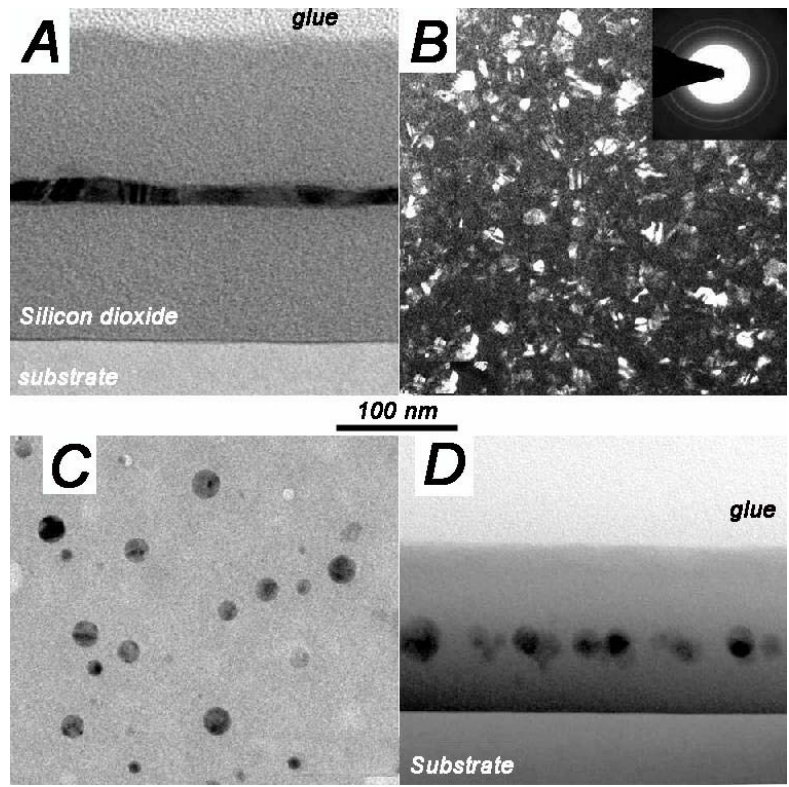

**FIGURE 2**  **a, b** Bright-field cross-section and dark-field in the Ge (111)-ring plan-view images of a sample from wafer, oxidized at 900 °C for 15 min in $O_2$. Continuous polycrystalline Ge layer is clearly observed; **c, d** Bright-field images of a plan-view and cross-section samples from wafer, oxidized at 900 °C for 30 min in $O_2$ followed by reduction at 900 °C for 30 s in $N_2$. Ge NCs show a dark contrast on the gray background of the $SiO_2$ layer.

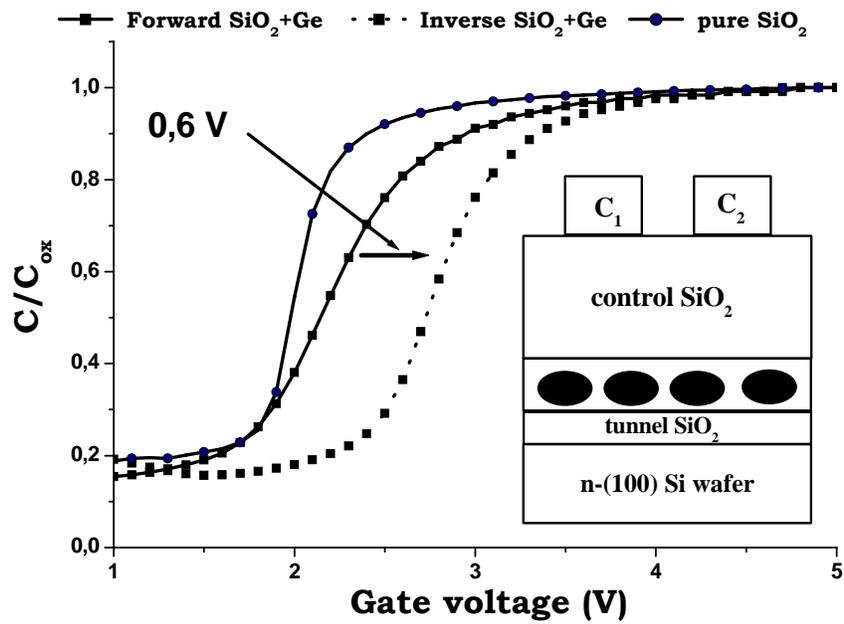

**FIGURE 3** High-frequency capacitance vs voltage curves of samples from wafer, oxidized at 900 °C for 30 min in dry $O_2$ followed by reduction at 950 °C for 30 s in $N_2$. High positive and negative gate voltages cause electron and hole injections respectively into the oxide from the substrate. A gate voltage ($Vg$) sweep from inversion to accumulation and from accumulation to inversion is shown on the figure by the forward and inverse directions respectively. The CV curve from reference sample of pure $SiO_2$ is also represented.

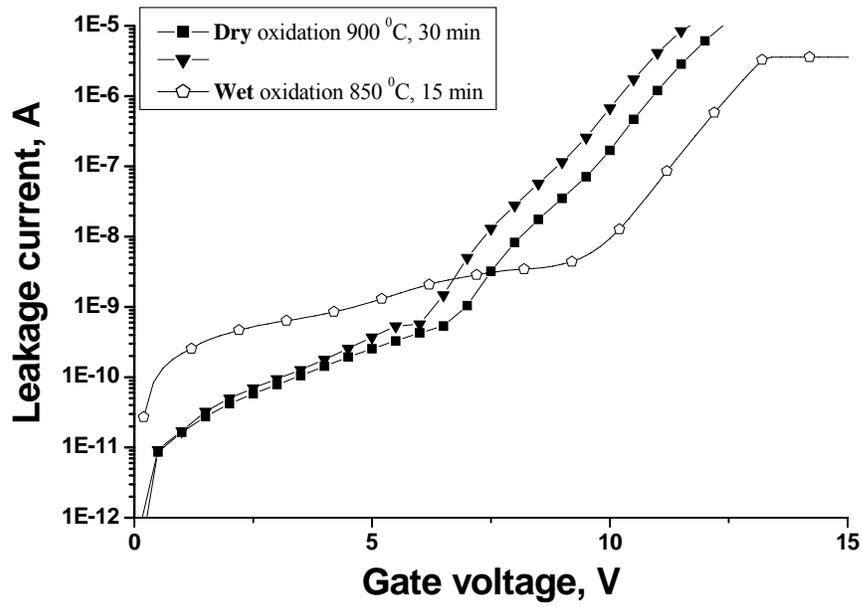

**FIGURE 4** Leakage current vs gate voltage characteriscics obtained from wafers oxidized in both wet and dry conditions at 850 °C and 900 °C corresponding.